# Dimensional crossover of the electronic structure in LaNiO$_3$ ultrathin films: Orbital reconstruction, Fermi surface nesting, and the origin of the metal-insulator transition

Hyang Keun Yoo,[1,2] Seung Ill Hyun,[3] Luca Moreschini,[4] Young Jun Chang,[4,5†] Da Woon Jeong,[1,2] Chang Hee Sohn,[1,2] Yong Su Kim,[1,2] Hyeong-Do Kim,[1,2] Aaron Bostwick,[4] Eli Rotenberg,[4] Ji Hoon Shim,[3] and Tae Won Noh[1,2]*

[1] Center for Functional Interfaces of Correlated Electron Systems, Institute for Basic Science, Seoul 151-747, Korea

[2] Department of Physics and Astronomy, Seoul National University, Seoul 151-747, Korea

[3] Department of Chemistry, Pohang University of Science and Technology, Pohang 790-784, Korea

[4] Advanced Light Source, Lawrence Berkeley National Laboratory, Berkeley, California 94720, USA

[5] Department of Physics, University of Seoul, Seoul 130-743, Korea

**Abstract**

Dimensionality control in the LaNiO$_3$ (LNO) heterostructure has attracted attention due to its two-dimensional (2D) electronic structure was predicted to have an orbital ordered insulating ground state, analogous to that of the parent compound of high-$T_c$ cuprate superconductors [P. Hansmann *et al.*, Phys. Rev. Lett. **103**, 016401 (2009)]. Here, we directly measured the electronic structure of LNO ultrathin films using *in situ* angle-resolved photoemission spectroscopy (ARPES). We recognized the dimensional crossover of the electronic structure around 3-unit cells (UC)-thick LNO film and observed the orbital reconstruction. However, complete orbital ordering was not achieved. Instead, we observed that the Fermi surface nesting effect became strong in the 2D LNO ultrathin film. These results indicated that the orbital reconstruction should be described by taking into account the strong nesting effect to search for the novel phenomena, such as superconductivity in 2D LNO heterostructure. In addition, the APRES spectra showed that the Fermi surface existed down to a 1-UC-thick film, which showed insulating behavior in transport measurements. We suggested that the metal-insulator transition in the transport properties may originate from Anderson localization.

Dimensionality is considered as one of the most important parameters to determine the physical properties in condensed matter physics [1,2]. Depending on the dimensionality of an electronic structure, the resulting physical properties can vary significantly [3-11]. For example, in the Ruddlesden-Popper series of perovskite structures, the dimensionality can be controlled from two- to three-dimensions, which induces a change in the electronic phase such as metal-insulator transition (MIT) [11]. Additionally, compared with three-dimensional (3D) metals, two-dimensional (2D) systems can have enhanced susceptibility near the Fermi surface (FS) or significant nesting, which results in novel quantum electronic phases, including density waves and high-$T_c$ superconductors [12-15]. Despite these interesting phenomena, there have been few studies on the dimensional crossover of the electronic structures in bulk materials due to experimental difficulties. However, recent advancements of the state-of-the-art film fabrication methods have allowed us to control the dimensionality of the electronic structure in diverse materials [16-20]. Moreover, combining with *in situ* angle-resolved photoemission spectroscopy (ARPES) system, we can look directly into the electronic structural changes in terms of the layer thickness [7,8].

The dimensional crossover of the electronic structure in LaNiO$_3$ (LNO) superlattices and ultrathin films is particularly interesting [16-25]. Recent theoretical calculations predicted that the 2D LNO system could have a fascinating electronic structure analogous to the parent compound of the cuprate superconductors [21-23]. Three-dimensional bulk LNO has a $3d^7$ electronic configuration with fully-occupied $t_{2g}$ and partially-filled degenerate $e_g$ orbitals [26]. Figure 1(a) shows the FS of the 3D metallic LNO, as calculated from density functional theory (DFT). There is a spherical FS centered at Γ point and a nearly cubic FS centered at A point, which have electron and hole band characteristics, respectively [26]. Figure 1(b) shows the DFT-calculated FS of the 2D LNO, such as a 1-unit cell (UC)-thick layer system. Note that the latter band structure has no dispersion along $k_z$, which indicates the 2D nature of the electronic bands [27]. When strong correlation energy was included, the theory predicted that 2D LNO under tensile strain could have significant $e_g$ orbital reconstruction and an insulating gap near the Fermi level ($E_F$) [21-23]. The predicted electronic structure just below $E_F$ is schematically shown in Fig. 1(c). Note that this is analogous to cuprate superconductors, i.e., a single $d_{x2-y2}$ orbital structure [22,28]. If realized, such an electronic structure of 2D LNO could provide another parent compound to search for superconductivity.

There are many experimental results that show the MIT associated with dimensional crossover in LNO heterostructures [16-20]. Many possible mechanisms have been proposed to explain it, such as the orbital ordered Mott insulator, charge disproportionation, density waves, and localization [16-25]. However, the origin of this insulating ground state is still controversial. In this letter, to clarify the nature of the 2D LNO system, we directly investigated the electronic structures of LNO films using *in situ* ARPES system integrated with the state-of-the-art film fabrication method. We reduced the LNO film thickness from 10 UC to 1 UC and observed that the dimensional crossover of the electronic structure occurred in around 3-UC-thick LNO film. This dimensional crossover could induce $e_g$ orbital reconstruction to push up the $d_{3z2-r2}$ orbital above $E_F$. However, complete $d_{x2-y2}$ orbital ordering, shown in Fig. 1(c), has not been achieved in our LNO ultrathin films. Instead, we recognized that the FS nesting effect became strong in 2D LNO ultrathin films, which induced band folding corresponding to a nesting vector. These results indicated that the orbital reconstruction should be described by considering the strong FS nesting effect to search for the novel phenomena, such as superconductivity in 2D LNO heterostructure. Furthermore, we observed a MIT in the transport properties between 3- and 4-UC-thick LNO films although the APRES spectra showed the existence of the FS down to a 1-UC-thick film. This indicated that the disorder effects, such as Anderson localization, could be important to understand the insulating state of the LNO ultrathin films. .

We prepared high-quality epitaxial $LaNiO_3/LaAlO_3$ films on (001)-oriented single crystal Nb:$SrTiO_3$ (+1.7% misfit tensile strain [17]) using pulsed laser deposition (PLD). We used a KrF excimer laser ($\lambda = 248$ nm) with a repetition rate of 2 Hz to ablate the sintered stoichiometric targets. The laser energy density was 1.0~1.5 J cm$^{-2}$ at the target position. Before PLD, atomically flat Nb-0.5 wt %-doped $SrTiO_3$ substrates were prepared by using buffered-hydrofluoric acid (HF) etching and heating processes. The $SrTiO_3$ substrates were conducting due to Nb doping, which prevented charging problems during the ARPES measurements. Additionally, to avoid possible intermixing between La and Sr, we deposited a 15-UC-thick LaAlO$_3$ buffer layer. Then, to control the dimensionality, we varied the LNO film thickness from 1 to 10 UC. In addition, to control for quality variation among films, we deposited all of the films between 1 and 10 UC at once by choosing the deposited area selectively using a mechanical shutter. The details of sample preparation are described in the Supplemental Material [29].

The *in situ* ARPES measurements were performed in an end station equipped with PLD and ARPES, at beam line 7.0.1 of the Advanced Light Source. After film deposition, the thin films were transferred in vacuum to the analysis chamber with a base pressure of 5×10$^{-11}$ Torr. The sample temperature was kept at around 90 K, except for specified cases. The total-energy resolution (photons + electrons) was around 30 meV, corresponding to a photon energy of 150 eV. We performed *in situ* ARPES measurements in Γ−X−M and Z−R−A symmetry planes [Fig. 1(d)]. To measure these symmetry planes, we selected appropriate photon energies for the ARPES measurements [29].

We investigated the dimensional crossover of the electronic structures in LNO films by varying the thickness from 1 to 10 UC. First, we observed that the FS of a 10-UC-thick LNO film had 3D character [Fig. 2(a)]. The electron (hole) band centered at Γ (A) point was consistent with the DFT-calculated 3D FS. Note that the additional hole band centered at M point may be the result of strong correlation effects, which are not included in the DFT calculations [29]. Then, we observed that the FS of the LNO films changed gradually from 10 to 1 UC, as shown in Figs. 2(a)–(e). Near Γ point, the size of the electron pocket became smaller as the film thickness decreased. The momentum distribution curves (MDCs) at $E_F$ along X−Γ−X showed the thickness-dependent change more clearly [Fig. 2(f)]. Namely, the electron pocket size remained nearly the same down to 4 UC. Then, it decreased gradually between 4 and 2 UC. On the other hand, around Z point, an electron pocket emerged as the thickness decreased. For the 10-UC-thick film, there was a low signal around Z [bottom panel of Fig. 2(a)]. However, the electron pocket became quite evident for samples having a thickness of 3UC or less. The MDCs at $E_F$ along R−Z−R in Fig. 2(g) show this feature more clearly. With these changes, the band at Z point became very similar to the one at Γ point, indicating that the LNO ultrathin films with 3 UC and below could have a quasi-2D FS with little dispersion along $k_z$.

The dimensional crossover can be more easily seen in the band dispersion curves in Fig. 3. There exists a clear change in the band dispersion along Γ−X−M for the 3-UC-thick sample [Figs. 3(a)–(e)]. This change was also observed along Z−R−A [Figs. 3(f)–(j)]. Furthermore, for the 10-UC-thick LNO film, the band dispersions along Γ−X−M and Z−R−A were clearly different [Figs. 3(a),(f)]. This difference indicates the 3D nature of the electronic structure. On the other hand, for the LNO films of 3 UC and below, the band dispersions along the high-symmetry points were nearly identical [Figs. 3(c)–(e) and (h)–(j)], suggesting that they have a quasi-2D electronic structure. Note that the FS crossing points along Γ−X and R−A changed gradually between 4 and 2 UC. The intensity at Z point was nearly low at 10 UC, increased around 4 UC, and began to form an electron-like band at 3 UC. All of these results support our argument that the dimensional crossover from 3D to quasi-2D electronic structure should occur in around 3-UC-thick LNO film.

Dimensional crossover can induce the orbital reconstruction, such as the upward shift of the $d_{3z^2-r^2}$ orbital, and suppression of the quasiparticle spectral weight (QSW). Figures 4(a),(b) display the electron band dispersions along X−Γ−X for 10- and 2-UC-thick films. This electron band is mainly composed of the $d_{3z^2-r^2}$ orbital in two dimensions [22,29]; thus, this band shifted upward as the thickness decreased. This shifting behavior can be more clearly seen in Fig. 4 (c). Additionally, figure 4 (d) shows the EDCs of the photoemission intensity at $k_F$ (A point) of the electron band and the strong suppression of the QSW is observed in the 2D regime. The main difference between 3D and 2D films is the hopping nature between the Ni $d_{3z^2-r^2}$ and O $p_z$ orbitals along the $z$-direction. For the 2D case, confinement along the $z$-axis can suppress such a hopping process, which induces the upward shift of the bottom of the $d_{3z^2-r^2}$ orbital band and suppression of the QSW [22,23]. However, such a complete single $d_{x^2-y^2}$ orbital structure and insulating gap opening, predicted by earlier theories [21-23], cannot be achieved in our LNO ultrathin films.

Although our APRES spectra showed that the FS exists down to 1-UC-thick film, there have been numerous reports on the insulating properties of such ultrathin LNO films [16-18] and superlattices [19-23]. To explain the insulating ground state, several mechanisms have been proposed, such as charge disproportionation and Mott insulator [19-23]. As shown in Fig. 4(e), our *ex situ* transport measurements also showed an insulating temperature dependence below 4 UC without an insulating gap opening. This lack of a band gap indicates that the observed insulating state cannot be explained either by charge disproportionation [19,20] or by the correlation-induced Mott gap [23]. Another possibility is that the insulating behavior can originate from generation of a large percolating cluster of insulating channels in ultrathin films [10]. If such a large cluster becomes generated, we should be able to observe the temperature dependent change in the photoemission spectra due to phase separation [32]. However, our ARPES data of 3 UC-thick LNO film was nearly the same in temperature [29]. Therefore, the percolation picture can be also ruled out as the origin of the insulating behavior in our LNO ultrathin films.

On the other hand, disorder effects, such as Anderson localization [16], could result in the insulating state for the ultrathin films with a FS. In two-dimensions, it is well known that the Anderson localization, caused by disorder effects, can occur more easily than in three-dimensions [30]. When the small density of states at the $E_F$ becomes significantly reduced, the mobility edges should move significantly away from the $E_F$ [31]. Then, the $E_F$ will be located between the mobility edges of the conduction and valence bands, so that the films can become insulating, despite the fact that the FS exists. The suppression of the QSW in Fig. 4 (d) can also be understood by considering the Anderson localization effect in 2D LNO ultrathin films as well as the suppression of hopping along the *z*-direction. This is consistent with the earlier transport study [16] that the Anderson localization should play an important role in the insulating state observed in quasi-2D LNO system.

We observed the clear change in the hole band in two-dimensions as shown in Fig. 4(f)–(i). Figures 4(f),(g) display the hole band dispersions along A–R–A for 10- and 2-UC-thick films, respectively. The hole band was mainly composed of the $d_{x2-y2}$ orbital [22,29] and became more dispersive in two-dimensions. This result indicates that the effective mass of holes occupying the $d_{x2-y2}$ orbital is reduced associated with dimensional crossover. This behavior is more clearly seen in Fig. 4(h). We speculate that the change of the $d_{x2-y2}$ orbital band may originate from the reduction of hopping between the $d_{x2-y2}$ and $d_{3z2-r2}$ orbitals in two-dimensions. Figure 4 (i) shows the EDCs obtained from the photoemission intensity at $k_F$ (B point) of the hole band. Interestingly, the QSW of the hole band was strongly suppressed in 2D regime although the effective mass of the hole was reduced. We normalized the EDCs of both the electron and hole bands with the oxygen peak located near −2.0 eV. Then, we fitted the EDCs using a Lorentzian function convoluted with a Gaussian function to estimate the change in the QSW [29]. The integrated QSWs of the electron and hole bands are plotted in Fig. 4 (j) as a function of film thickness. We observed that the integrated QSW of the hole band decreased faster than that of the electron band in two-dimensions. If the suppression of hopping along the *z*-direction and Anderson localization are the only effects induced by dimensional crossover in LNO films, the integrated QSW of the electron band ($d_{3z2-r2}$ orbital) should be more suppressed than that of the hole band ($d_{x2-y2}$ orbital). Therefore, this result indicates that there should be an additional effect induced by dimensional crossover.

We observed that the FS nesting effect became strong in the 2D LNO ultrathin film, which induced a band folding corresponding to the nesting vector. The shape of the FS became very flat in two-dimensions as shown in Fig. 2 and Fig. 5(a). These large flat regions and itinerant property of carriers in the hole band in 2D LNO ultrathin film can induce a strong FS nesting effect. The nesting vector **q** is given by $\frac{\pi}{a}(\frac{1}{2},\frac{1}{2},0)$ indicated by blue arrows in Fig. 5(a) [13,24]. The value of **q** was found to be close to the wave vector for spin-density-wave (SDW) observed in all of the insulating nickelates [24]. With this nesting vector, an additional folded band should appear. A description of the reconstructed FS due to this SDW is given in Fig. 5(b). The experimentally measured FS folding pattern in Fig. 5(a) is similar to that shown in Fig. 5(b). Although we did not observe the clear band gap opening, we expected that the large FS nesting enhanced the susceptibilities at the nesting vector **q** and produced an additional folded band with SDW fluctuation in the 2D LNO ultrathin film [24]. Note that similar band folding due to SDW has been observed in other correlated electron systems such as unconventional superconductors [33,34]. The strong suppression of QSW in Fig. 4(j) was also attributed to the strong spectral weight transfer from the hole band to the additional folded band. This result suggests that we should consider not only the orbital reconstruction but also the strong FS nesting effect when we design the novel phenomena such as superconductivity in LNO heterostructure.

In summary, we directly investigated the dimensional crossover by visualizing the electronic structure of LNO ultrathin films, combining *in situ* APRES system with the state-of-the-art film fabrication methods. Our investigation of LNO ultrathin films demonstrates how dimensional crossover of an electronic structure evolves in correlated electron systems. Furthermore, we were able to clarify controversial issues such as orbital reconstruction, FS nesting effect, and the origin of the MITs in transport measurement in 2D LNO system. We expect that our experimental results can serve as new platform to search for novel phenomena in other correlated electron systems using *in situ* ARPES system..

This work was supported by the Research Center Program of IBS (Institute for Basic Science) in Korea (Grant No. EM1203). J.H.S. was supported by the National Research Foundation of Korea (NRF) (2012029709). L.M. acknowledges support by a grant from the Swiss National Science Foundation (SNSF) (project PBELP2-125484). Y.J.C. acknowledges support from National Research Foundation of Korea under Grant No. NRF-2012R1A1A2043619. The Advanced Light Source is supported by the Director, Office of Science, Office of Basic Energy Sciences, of the U.S. Department of Energy under Contract No. DE-AC02-05CH11231. The 9C beam line in Pohang Accelerator Laboratory is used for this study.

†e-mail: yjchang@uos.ac.kr

*e-mail: twnoh@snu.ac.kr

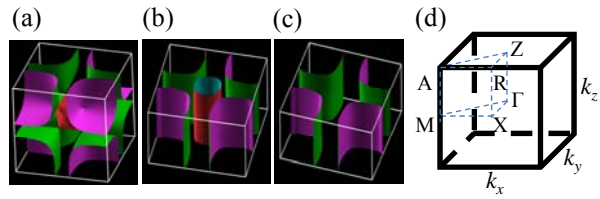

Fig. 1. (color online). Density functional theory (DFT) calculations for the Fermi surfaces (FSs) of (a) the bulk LaNiO$_3$ (LNO) and (b) two-dimensional (2D) LNO. (c) A schematic picture of the predicted electronic structure just below the Fermi level ($E_F$) of 2D LNO when strong correlation effects were taken into account. (d) The symbols for the high-symmetry momentum points (Γ, X, M and Z, R, A) come from the Brillouin zone of the tetragonal structure.

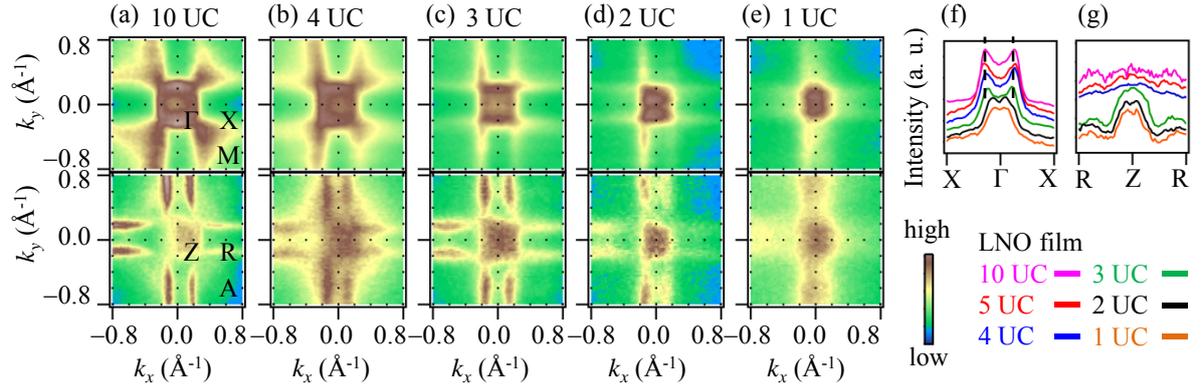

Fig. 2. (color online). (a)–(e) FSs of two symmetric planes (Γ−X−M and Z−R−A) for 1- to 10-unit cells (UC)-thick LNO films under +1.7 % tensile misfit strain comes from the SrTiO$_3$ substrate. The electronic structures in Γ−X−M and Z−R−A planes became nearly identical in the LNO films thinner than the 4-UC-thick LNO film. Momentum distribution curves (MDCs) along (f) X−Γ−X and (g) R−Z−R at the $E_F$. These MDCs were obtained with an energy resolution of 50 meV and clearly show the dimensional crossover of the electronic structure around 3 UC. The dashed lines in (f) are guidelines indicating peak positions in the MDCs.

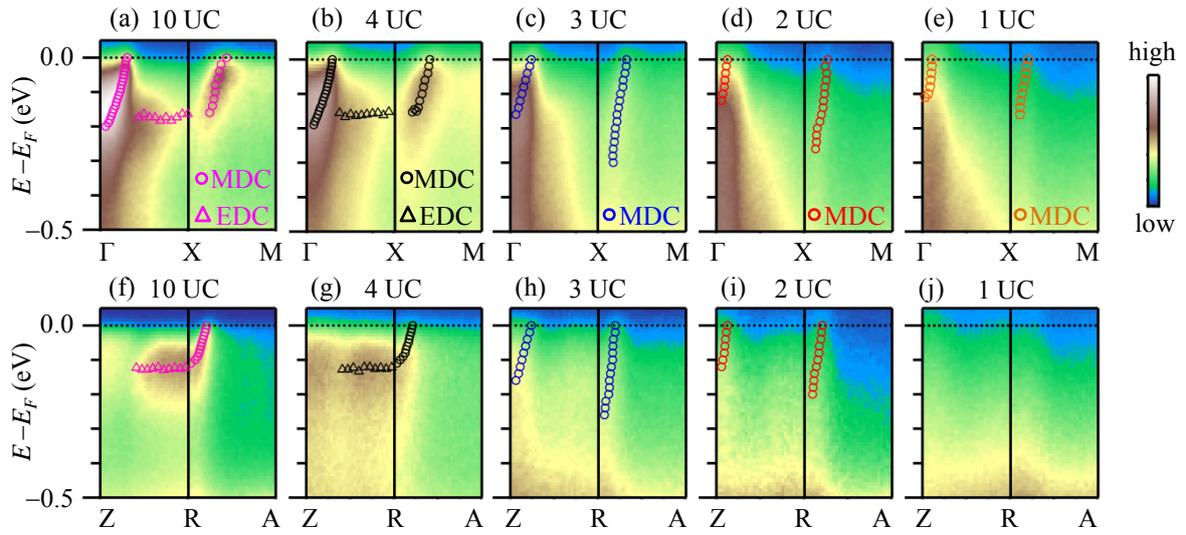

Fig. 3. (color online). (a)–(j) Electronic band dispersions along Γ−X−M and Z−R−A points for 1- to 10-UC-thick LNO films. We obtained the band dispersions along Γ−X−M and Z−R−A by tracking the peak positions experimentally in the MDCs (empty-circles) and energy distribution curves (EDCs, empty-triangles). Along Z−R−A in (j), the MDC and EDC signals of the 1-UC-thick LNO film were too weak to determine the corresponding peak positions.

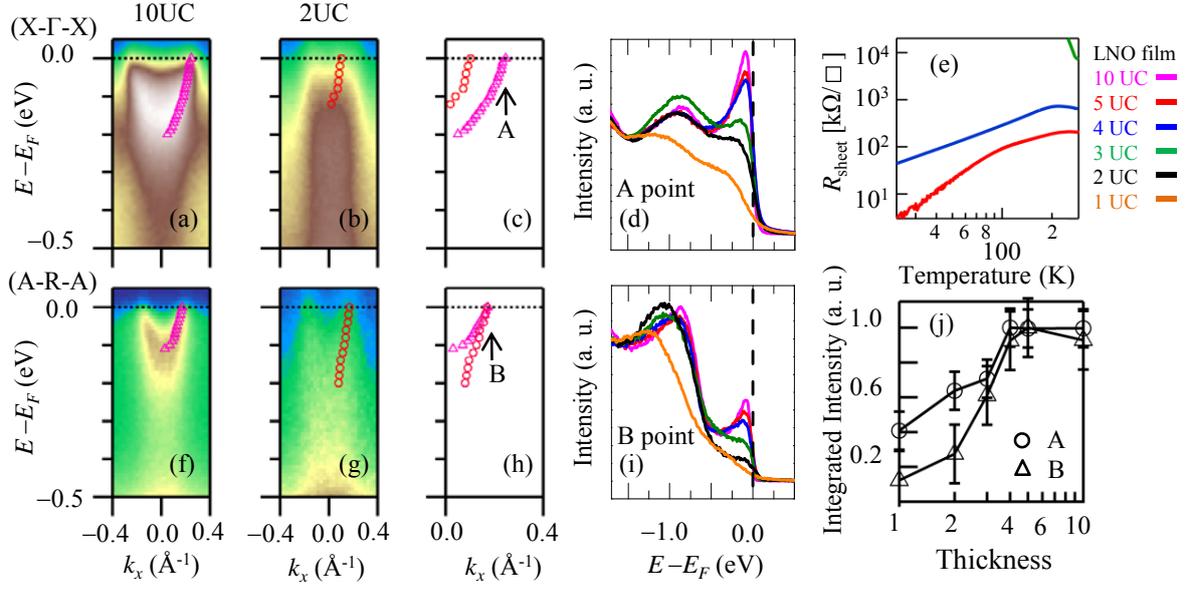

Fig. 4. (color online). (a)–(d) The changes of the electron band dispersions along X–Γ–X for 10-, 2-UC-thick films and obtained EDCs at $k_F$ (A point) associated with dimensional crossover. The strong suppression of the quasiparticle spectral weight (QSW) is observed in the 2D regime. (e) The insulating temperature dependence in transport measurement was observed although the photoemission spectra show that the FS existed down to a 1-UC-thick film. (f)–(i) The changes of the hole band dispersions along A−R−A for 10-, 2-UC-thick films and obtained EDCs at $k_F$ (B point) associated with dimensional crossover. (j) We found that the integrated QSW of the hole band decreased more than that of the electron band in two-dimensions. Error bars are determined from the sizes of background signals.

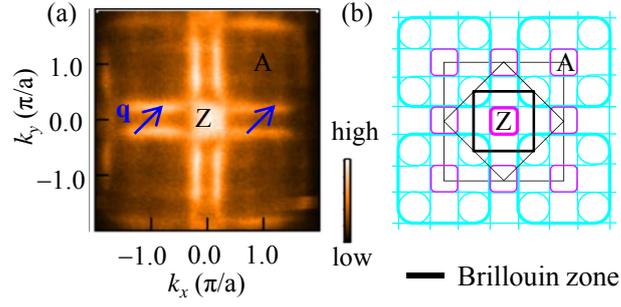

Fig. 5. (color online). (a) The FS of 3-UC-thick LNO in Z−R−A symmetry plane. The shape of the FS became very flat in two-dimensions, which could induce a strong FS nesting effect. The nesting vector **q** was given by $\frac{\pi}{a}(\frac{1}{2},\frac{1}{2},0)$ indicated by blue arrows and additional folded band appeared. The found nesting vector **q** was close to the wave vector for spin-density-wave (SDW) observed in all of the insulating nickelates. A description of the reconstructed FS due to this SDW was shown in (b).